\def\year{2022}\relax
\documentclass[letterpaper]{article} % DO NOT CHANGE THIS
\usepackage{aaai22}  % DO NOT CHANGE THIS
\usepackage{times}  % DO NOT CHANGE THIS
\usepackage{helvet}  % DO NOT CHANGE THIS
\usepackage{courier}  % DO NOT CHANGE THIS

\usepackage{soul}
\usepackage[hyphens]{url}  % DO NOT CHANGE THIS
\usepackage{graphicx} % DO NOT CHANGE THIS
\usepackage{natbib}  % DO NOT CHANGE THIS AND DO NOT ADD ANY OPTIONS TO IT
\usepackage{caption} % DO NOT CHANGE THIS AND DO NOT ADD ANY OPTIONS TO IT
\DeclareCaptionStyle{ruled}{labelfont=normalfont,labelsep=colon,strut=off} % DO NOT CHANGE THIS
\usepackage{algorithm}
\usepackage{algorithmic}
\usepackage{enumitem}

\usepackage{xcolor}
\newcommand{\answerYes}[1]{\textcolor{blue}{#1}} 
\newcommand{\answerNo}[1]{\textcolor{teal}{#1}} 
\newcommand{\answerNA}[1]{\textcolor{gray}{#1}} 
 
\usepackage{newfloat}
\usepackage{listings}
\floatstyle{ruled}
\newfloat{listing}{tb}{lst}{}
\floatname{listing}{Listing}
    \makeatletter
\def\@fnsymbol#1{\ensuremath{\ifcase#1\or \dagger\or \ddagger\or
   \mathsection\or \mathparagraph\or \|\or **\or \dagger\dagger
   \or \ddagger\ddagger \else\@ctrerr\fi}}
    \makeatother

\title{Toxic Synergy Between Hate Speech and Fake News Exposure}

\author{
   Munjung Kim$^{1}$, Tuğrulcan Elmas$^{2}$, Filippo Menczer$^{1}$}
\affiliations{

$^1$ Indiana University Bloomington

 $^{2}$University of Edinburgh

munjkim@iu.edu, telmas@iu.edu, fil@indiana.edu 
}

\begin{document}

\maketitle

\begin{abstract}
Hate speech on social media is a pressing concern. Understanding the factors associated with hate speech may help mitigate it. 
Here we explore the association between hate speech and exposure to fake news by studying the correlation between exposure to news from low-credibility sources through following connections and the use of hate speech on Twitter. 
Using news source credibility labels and a dataset of posts with hate speech targeting various populations, we find that hate speakers are exposed to lower percentages of posts linking to credible news sources. 
When taking the target population into account, we find that this association is mainly driven by anti-semitic and anti-Muslim content. 
We also observe that hate speakers are more likely to be exposed to low-credibility news with low popularity. 
Finally, while hate speech is associated with low-credibility news from partisan sources, we find that those sources tend to skew to the political left for antisemitic content and to the political right for hate speech targeting Muslim and Latino populations.
Our results suggest that mitigating fake news and hate speech may have synergistic effects. 
\end{abstract}

\section{Introduction}

Hate speech is a barrier to inclusivity and harmonious coexistence in a diverse society. 
The proliferation of misinformation, on the other hand, challenges healthy public dialogue. 
Although both issues are harmful and should be mitigated on their own, we conjecture that they may also synergistically amplify each other. 
In this paper, we explore this possible link by asking: \textit{``Is the use of hate speech associated with exposure to fake news?''} We attempt to answer this question on Twitter (later renamed X) by studying two groups of users, \textit{hate speakers} who use hate speech on social media toward certain target populations and \textit{non-hate speakers} who also speak about the same populations but in non-hateful ways. We start from a preexisting dataset in which posts are annotated to contain hate speech or not. We expand this data by (i)~identifying the authors of those posts, (ii)~collecting the accounts followed by them and their tweets, (iii)~identifying the news articles shared in those tweets, and finally (iv)~using the credibility of the news sources to measure fake news exposure. 
%We then compared the two groups. 
We focus on the following research questions:
\begin{enumerate}[label=\textbf{{RQ{\theenumi}}},leftmargin=26pt]
    \item Are hate speakers exposed to more fake news through their following connections?
\end{enumerate}
Is the relationship between hate speech and fake news exposure affected by...
\begin{enumerate}[resume,label=\textbf{{RQ{\theenumi}}},leftmargin=26pt]
    \item ... the groups targeted by the hate speakers? 
    \item ... the popularity of the fake news? 
    \item ... the political orientation of the fake news? 
\end{enumerate}

In answering these questions, we found that hate speakers are exposed to a higher proportion of news from low-credibility sources through the accounts they follow when compared to non-hate speakers. 
This finding holds for hate speech targeting Jews and Muslims; differences for other target groups are not significant.
We also found that the difference in exposure to news from low-credibility sources was mainly due to unpopular tweets linking to far-right sources for users targeting Muslims and far-left sources for those targeting Jews.

\section{Related Work}

% Related work is good for a long paper but maybe too long for a short one. 
% The study of features of hate speakers engaging in hate speech or the spread of hate speech has been a global focus. Hate speakers often maintain anonymity, display infrequent account verification and geotagging, and exhibit higher follower and following counts, favorites, group memberships, and statuses~\cite{perera2023comparative}. 

% While our work is the first to study the association between hate speech and fake news, both problems have been studied extensively.

%\noindent\textbf{Exposure to Fake News:}
\paragraph{Exposure to Fake News.} Fake news pose research challenges to both social science and computer science~\cite{lazer2018science}. Previous work tackled their detection by analyzing their spread or exposure. For instance, \citet{grinberg2019fake} found that the consumption of fake news was highly concentrated among users and that user age and the number of political links in the news feed were strong predictors of exposure to fake news. \citet{bovet2019influence} reported that during the 2016 U.S. elections, the activity of Clinton supporters was influenced by top influencers spreading traditional news, while Trump supporters influenced the dynamics of the top fake news spreaders. Both studies underline that certain individuals (e.g., older and more conservative voters) are more vulnerable to fake news. We expand upon these studies by showing that individuals who use hate speech are also more susceptible to fake news.

%\noindent\textbf{Use of Hate Speech:} 
\paragraph{Use of Hate Speech.} 
%Past research analyzed the characteristics of hate speakers. 
\citet{perera2023comparative} reported that hate speakers often maintain anonymity, display infrequent account verification and geotagging, and tend to have more followers/following, favorites, group memberships, and statuses than non-hate speakers. Despite these characteristics, engagement with hate content from general Twitter users is low. \citet{cinelli2021dynamics} found that ``serial haters" ---active users posting exclusively hateful comments--- are uncommon. \citet{mathew2019spread} reported that content by hate speakers spreads faster and reaches a broader audience. Hate speakers also exhibit denser connections among themselves, emphasizing the cohesive nature of hate-driven online communities. 

\paragraph{Association Between Hate Speech and Fake News.} 
Recently, \citet{mosleh2024misinformation} delved into the use of generally harmful language among Twitter users who share misinformation. Posts containing links to lower-quality news outlets tend to contain more harmful language, and false headlines are more likely to include harmful language compared to true headlines. Moreover, Twitter users who share links to low-quality news sources also tend to use more harmful language. Rather than analyzing harmful language by people who share misinformation, our present investigation focuses on exposure to misinformation by users who produce hate speech. Following someone means subscribing to their updates or posts on social media platforms, indicating a level of interest or affinity with the content they produce. Therefore, studying the posts generated by the users followed by the hate speakers can elucidate the type of information that the hate speakers choose to engage with.
%We extend these studies by analyzing the hate speakers' exposure to fake news.
% ~\cite{schafer2018online}

\section{Data and Methodology}

% Short introduction to the section
We aim to study the correlation between the use of hate speech and exposure to low-credibility news. To this end, we collect three datasets from Twitter: (i)~users who do or do not produce hate speech, (ii)~posts by the accounts followed by those users, and (iii)~the credibility of those posts. 

% \subsection{Collecting hate speakers}
% \begin{figure}[h]
%     \centering
%     \includegraphics[width=0.4\textwidth]{Figure/keyword_distributions.png}
%     \caption{Caption}
%     \label{fig:keyword_distribution}
% \end{figure}

\paragraph{User Dataset.} 
While hate speech encompasses many forms of discrimination including gender, sexual orientation, and disability among others, our analysis primarily focuses on racism-related hate speech due to the opportunity presented by the availability of two existing annotated hate speech datasets. 
The first dataset comprises 5,880 tweets including the keywords ``Asians,'' ``Blacks,'' ``Jews,'' ``Latinos,'' and ``Muslims.'' These tweets were annotated by 120 human annotators. Each tweet was annotated by 3--7 people, depending on the keyword. As a result, 357 tweets (6.1\%) were labeled as hate speech against minority groups~\cite{jikeli_gunther_2023_7932888}. In cases of disagreement, the labels were selected based on the majority of the annotators. 
The second dataset comprises 6,941 tweets related to Jews including derogatory terms,\footnote{We do not list the terms here as they are offensive.} 
%``Kikes,'' ``Jews,'' ``ZioNazi,'' and ``Israel,'' 
spanning the period from January 2019 to December 2021. Each tweet was annotated by two experts, out of a pool of 10 different experts. Among these tweets, 1,250 (18\%) were classified as hate speech, adhering to the IHRA definition of antisemitic content~\cite{jikeli_gunther_2023_8147308}. In cases of disagreement, the annotators reached a consensus through discussion. 
After combining these two datasets, we extracted two sets of users: one that mentions hate speech and one that does not, despite discussing the same topic. 
Specifically, we randomly sampled 630 users who authored at least one tweet annotated as hate speech. We label these users as \textit{hate speakers}. 444 of them targeted Jews, 68 targeted Blacks, 74 targeted Muslims, 23 targeted Latinos, and 21 targeted Asians. 
We also sampled from this dataset 630 users who did not post hate speech as the control group, labeling them as \textit{non-hate speakers}. 502 of them mentioned keywords about Jews, 35 about Blacks, 34 about Muslims, 34 about Latinos, and 25 about Asians. 

\begin{figure}
    \centering
    \includegraphics[width=\columnwidth]{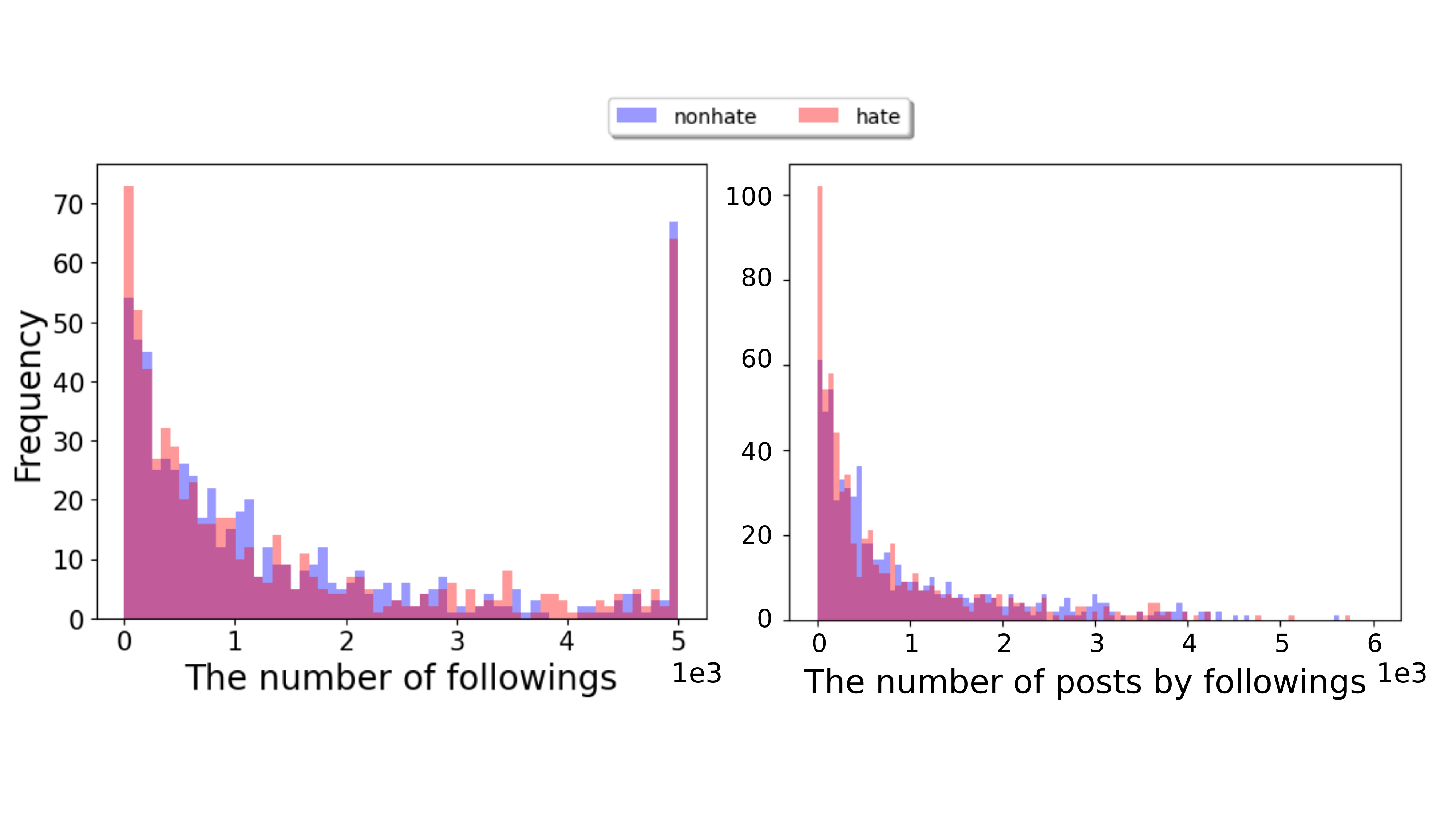}
    \caption{The distributions of the number of followings (left) and posts by followings (right) of hateful and non-hate speakers. The peak in the distribution of the number of followings at 5,000 is attributed to our imposed limitation, restricting the number of followings to a maximum of 5,000.}
    \label{fig:followees_distribution}
\end{figure}

% \subsection{Tweets Exposed to Users}

% \begin{figure}
%     \centering
%     \includegraphics[width=0.45\textwidth]{Figure/With_outlier/Figure1_followers_and_posts.pdf}
%     \caption{The distribution of the number of users followed by hate and non-hate speakers (left) and their posts (right). We collected up to 5,000 followings, resulting in the peak in the left distribution.}
%     \label{fig:followees_distribution}
% \end{figure}

%Collecting Tweets Exposed to hate speakers
\paragraph{Exposure Dataset.} We assume that the posts by the users followed by hate and non-hate speakers are their main sources of exposure to low-credibility news. As such, we use those posts as a proxy for exposure. Thus, we first collected the accounts that are followed by the users in our dataset. 
%To maintain data consistency and exclude outliers, 
Due to Twitter API limits, we were able to collect up to the most recent 5,000 followings for both hate and non-hate speakers (52 hate speakers and 53 non-hate speakers had more than 5,000 followings). 
We then collected the tweets that these followed accounts had authored in the previous 100 days. 
For each hate/non-hate speaker, the 100-day period preceded their post (or, if they posted more than one hate/non-hate tweet, it preceded one of these posts selected at random). 
As a result, we collected 48,637,629 posts from 943,671 users followed by 630 hate speakers and 55,435,787 posts from 960,752 users followed by 630 non-hate speakers.  The distributions of the number of users followed by each hate speaker and the number of posts produced by those users are shown in Figure.~\ref{fig:followees_distribution}. 
%The distributions of the number of followings and the number of posts by followings of each user used in the analysis are shown in Fig.~\ref{fig:followees_distribution}. 

% Among the 630 "hate speakers," 40 had less than 10 tweets posted by their followings, and among the 630 "non-hate speakers," 35 had less than 10 tweets posted by their followings. We excluded these users from our analysis.

\paragraph{Post Credibility.} As it is challenging to evaluate the credibility of each post with high accuracy, we focus on posts that link to news sources. We use source credibility as a proxy for post credibility. We use source credibility scores provided by NewsGuard, which evaluates news and websites using nine journalistic criteria. These criteria are converted into a \textit{trust score} ranging from 0 to 100. Sources with score below 60 are considered unreliable because they fail to adhere to several basic journalistic standards~\cite{Newsgaurd_standard}. In this paper, we classify these sources as low-credibility. We also use the political alignment (far-left, slight-left, slightly-right, far-right) of the sources, also provided by NewsGuard. Among our 104,073,416 posts, 3,512,022 (3.4\%) linked to one of 5,224 distinct sources annotated by NewsGuard. We focus on users exposed to at least 10 tweets linking to these sources, excluding 40 hate and 35 non-hate speakers. 

%also used the same criteria, which are ``trust score" of 40 and 60, to classify the low credible news. 

% Next, we classified the uncredible news by using the "trust score" provided by NewsGuard. Generally, the trust score below 40 is considered as "the website that is unreliable because it severely violates basic journalistic standards"~\cite{Newsgaurd_standard}.

\section{Results}

% \subsection{Ratio of Uncredible News}

% We first examined whether hate speakers who speak hate speech are more exposed to uncredible posts than non hate speakers by comparing the proportion of the posts including uncredible domain within the posts exposed to each hateful user with non-hateful user.  We classified the uncredible news by using the "trust score" provided by NewsGuard. Generally, the trust score below 40 is considered as "the website that is unreliable because it severely violates basic journalistic standards"~\cite{Newsgaurd_standard}. We also considered the posts including the domain with score below 40 as uncredible news. 

% Fig.~\ref{fig:proportion_uncredible} shows the result. 

%\subsection{RQ1: Are hate speakers exposed to the low-credible news more than non-hate speakers? }

\subsection{RQ1: Exposure of Hate vs. Non-Hate Speakers}

% \begin{figure}[h]
%     \centering
%     \includegraphics[width=0.5\textwidth]{Figure/Figure2_ccdf_exposure_uncredible.pdf}
%     \caption{CCDF }
%     \label{fig:ratio_low_credible}
% \end{figure}

% {Figure/With_outlier/Notlogscale_Ratio_absolute_median_credible_news.pdf} ==> without text

\begin{figure}
    \centering
    \includegraphics[width=\columnwidth]{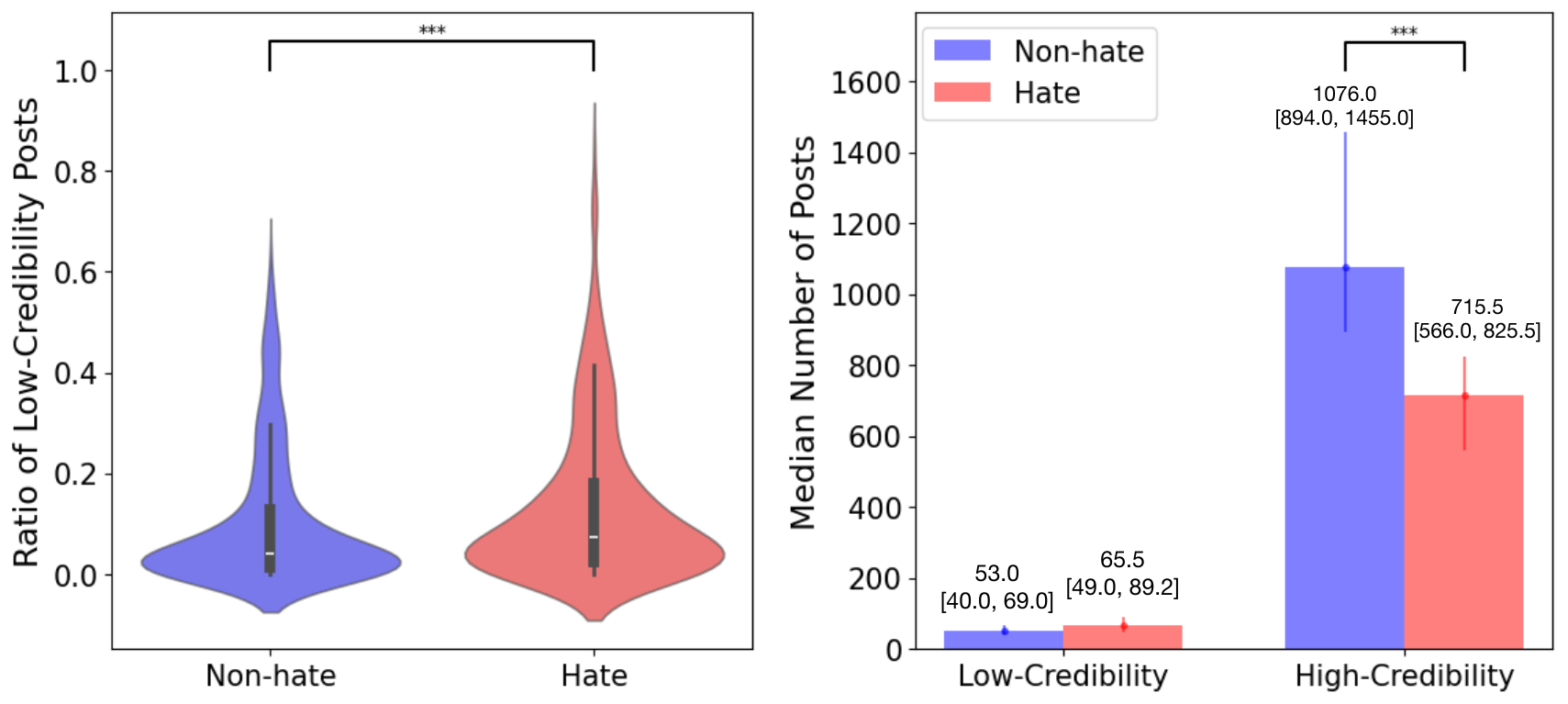}
    \caption{Left: Distributions of the ratio of low-credibility posts to which hate and non-hate speakers are exposed. Right: Median numbers of posts linking to low-credibility and credible news sources to which hate and non-hate speakers are exposed. Error bars indicate 95\% Bias-Corrected and Accelerated (BCa) confidence interval~\cite{efron1987better}. Significant differences in this and the following figures are indicated (***: $p<0.001$, **: $p<0.01$, *: $p<0.05$).}
    \label{fig:ratio_low_credible}
\end{figure}

% absolute exposure with threshold

We first investigate whether hate speakers are exposed to more fake news compared to non-hate (control) speakers. For each user, we collect the tweets to which they are exposed and compute the proportion linking to low-credibility sources. The median proportion of low-credibility posts is 0.073 for hate speakers and 0.040 for the control group. 
The distributions are shown by the violin plots in Fig.~\ref{fig:ratio_low_credible}. To statistically compare the proportions of low-credibility tweets to which hate and non-hate speakers are exposed, we use two-tailed Mann-Whitney U tests since the distributions are not normal. 
%The Mann-Whitney U test is a non-parametric statistical test used to evaluate the significance difference between two independent groups~\cite{fay2010wilcoxon}. It assigns ranks to all data points from two groups and compares the sums of the ranks for each group to determine the statistical difference between the two groups. 
%The result of the test reveals a significant difference between the proportions. 
The proportion for hate speakers is significantly higher than for non-hate speakers 
%($U1 = 204421$)  ($U2 = 146629$). 
($p < 0.001$).

%\textbf{What does this linktells us in a human readable way?}

% The result was consistent when the threshold is 60 ($U1 = 220281, U2 = 173473, p< 0.001$). 

% The denominator is all the posts containing the urls in the next paragraph

%We showed that the proportion of tweets with low credible news among all the tweets exposed to hate speakers is significantly higher than the tweets exposed to non-hate speakers. However, this result can be due to the case where hate speakers are less exposed to tweets with news domains regardless of their credibility. Therefore, we further investigate the proportion of the tweets with low credible news domains among only the tweets with news domains. We find that the result does not change much (Right figure in Fig.~\ref{fig:ratio_low_credible}) -- hateful users are exposed to a higher proportion of tweets with low credible news within the tweets including news domain than non-hateful users. 

% \begin{figure}[h]
%     \centering
%     \includegraphics[width=0.5\textwidth]{Figure/Ratio_low_credible_news_among_news.png}
%     \caption{CCDF }
%     \label{fig:proportion_all_news}
% \end{figure}

% absolute number

The proportion of low-credibility news masks the differences in the actual exposure values. Therefore, let us compare the numbers of posts linking to news articles from sources with low and high credibility to which users in the two groups are exposed. The median numbers of posts are also shown in Fig.~\ref{fig:ratio_low_credible}. 
We find that the number of posts linking to low-credibility sources for hate speakers is not significantly higher than for non-hate speakers ($p=0.36$). However, a significant difference exists when we compare the numbers of posts that link to credible news sources ($p<0.001$). These results suggest that hate and non-hate speakers are exposed to similar (small) volumes of low-credibility news but hate speakers are exposed to fewer posts linking to credible news sources.

\subsection{RQ2: Hate Speakers with Different Targets}

\begin{figure}[t!]
\includegraphics[width=\columnwidth]{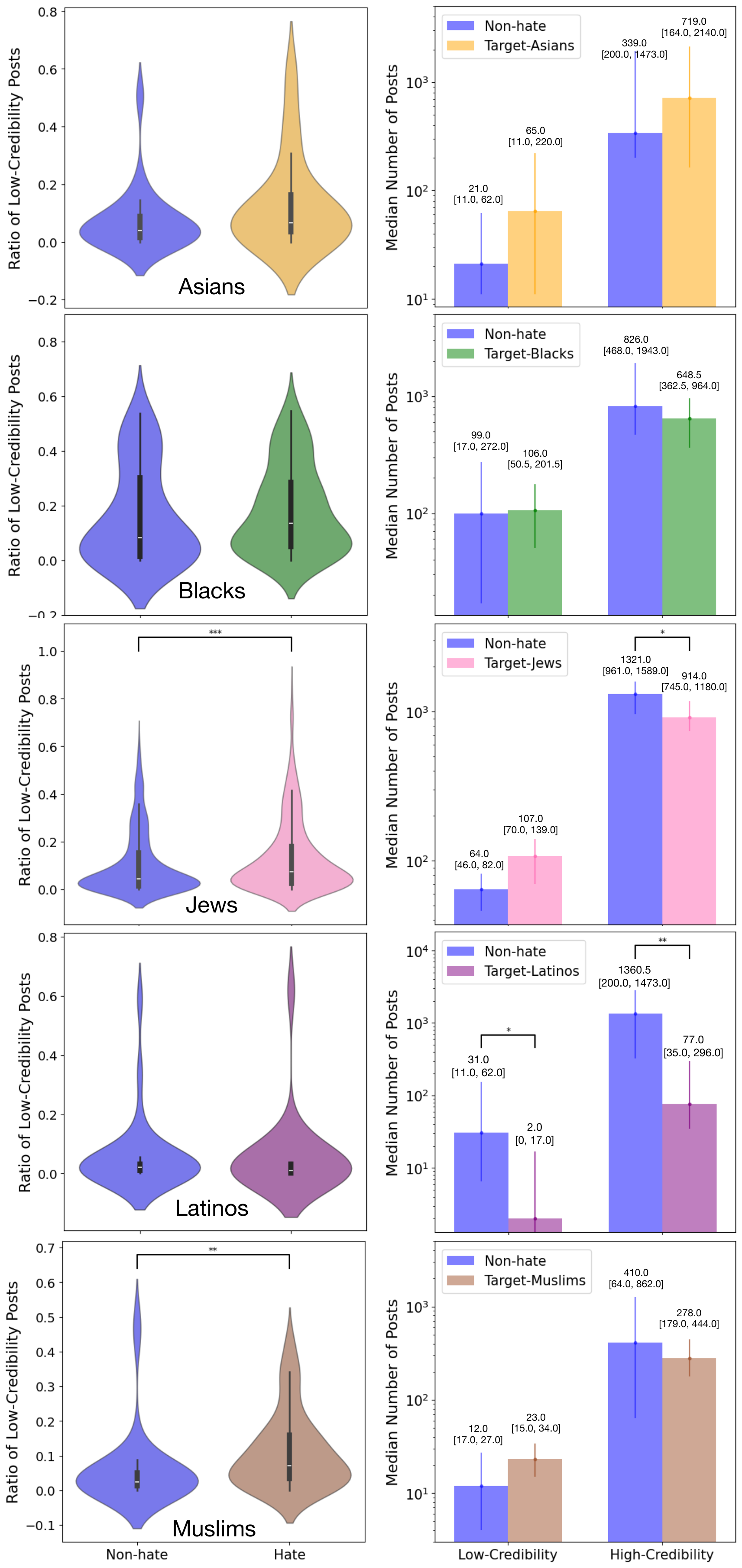}\centering
  \caption{Left: Proportions of posts linking to low-credibility news sources to which hate and non-hate speakers are exposed, for each population targeted by hate speech. Right: Median numbers of posts linking to low-credibility and credible news sources to which hate and non-hate speakers are exposed, for each target population.} \label{fig:absolute_number_news_total_value_race}
\end{figure}

% Here, we test if there are key differences between the users who target different groups, i.e., different kind of hate groups... \textbf{Do we already have some analyses here?}

We reported that hate speakers are exposed to a higher proportion of news from low-credibility sources. In this section, let us test if this result holds for hate speakers targeting different populations. 
Fig.~\ref{fig:absolute_number_news_total_value_race} presents the distributions for each target population. 
%violin plots showing the ratio of news with low credibility exposed to hate speakers targeting different population groups and the number of posts containing credible news exposed to the non-hate speakers and hate speakers targeting each group. 
%Using the two-tailed Mann-Whitney U test, we investigate a difference in the exposure to news with low credibility between non-hate speakers and hate speakers targeting specific population groups. 
We compare hate and non-hate speakers who posted tweets including the same population keywords (e.g., ``Asians'') to make the control group contextually relevant in each case. 
Mann-Whitney U tests show that only hate speakers targeted Jews ($p<0.001$) and Muslims ($p<0.01$) are exposed to a significantly higher proportion of low-credibility tweets than the corresponding non-hate speakers. 
This suggests that the pattern observed across all hate speech is mainly driven by data about anti-semitic and anti-Muslim content; this may be partly explained by the predominance of anti-semitic content in our dataset.  

Next, we compare the raw numbers of posts with low- and high-credibility news to which users are exposed for each target population. The results are also shown in Fig.~\ref{fig:absolute_number_news_total_value_race}. We observe that hate speakers targeting Jews are exposed to significantly fewer credible posts than non-hate speakers ($p = 0.036$). Hate speakers targeting Latinos have significantly lower exposure to credible tweets ($p<0.01$) than non-hate speakers, but also significantly lower exposure to low-credibility tweets ($p = 0.02$). 
We do not observe significant differences for the remaining target populations; this is possibly due to the sparsity of annotated hate speech targeting those populations in our dataset. 

%%%%% non categorized text %%%%%%%%%%%%%%%%

%%%%%%%%%%%%%%%%%%%%%%%%%%%%%%%%%%%%%%%%%%%%
%In conclusion, the ratio of low credible news in tweets exposed to hate speakers was higher than the non-hate speakers except for those targeting Asians. Also, the average ``trust scores'' of the news domains exposed to hate speakers targeting Blacks and Jews were significantly lower than non-hate speakers according to the independent sample t-test. 

% asian: the ratio of low credible news is not higher than non hate speakers. 
% muslims: the ratio of low credible news is higher than non hate speakers but the number of credible news and low credible news are both lower than non-hate speakers
% Jews: the ratio of low credible news is higher than non hate speakers and but this is due to the low credible news. 
% Black: the ratio of low credible news is higher than non hate speakers and this is because the number of credible news is lower. 

\subsection{RQ3: Exposure to Popular Fake News}

%no text==> {Figure/With_outlier/Engagement_vs_Proportion...}

\begin{figure}[t!]
\includegraphics[width=\columnwidth]{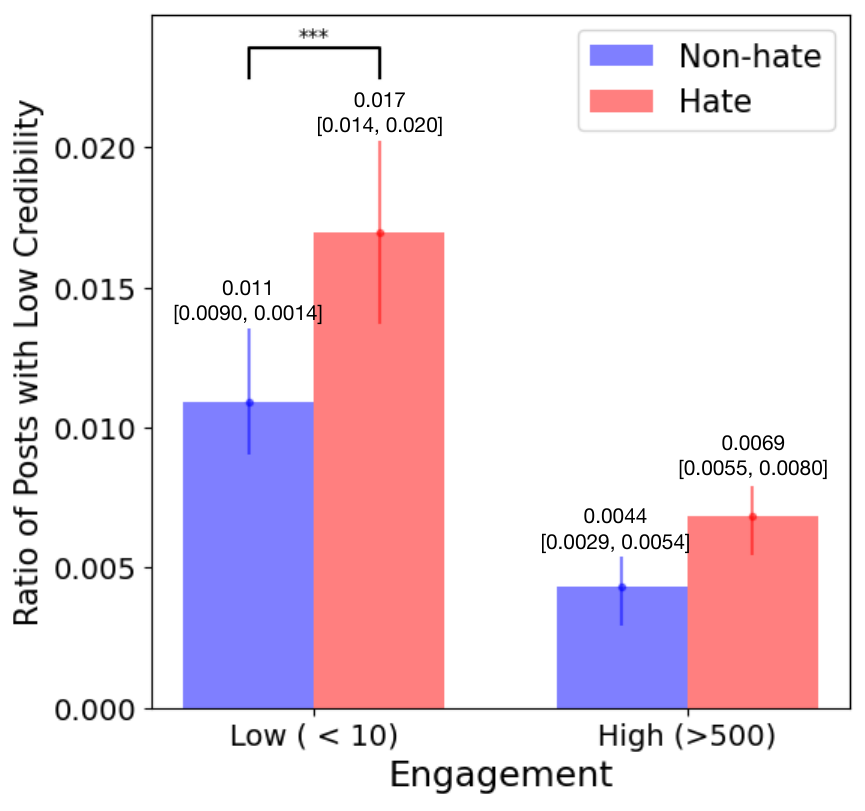}\centering
  \caption{Median proportions of tweets linking to low-credibility news sources to which hate and non-hate speakers are exposed, broken down by popularity: less than 10 vs. more than 500 likes and retweets.}
  \label{fig:engagement}
\end{figure}

% What is the nominator and denominator here.

We next investigate the popularity of the low-credibility news to which hate speakers are exposed. Let us define \textit{popular posts} as low-credibility tweets with more than 500 engagements (likes + retweets) and \textit{unpopular posts} as those with less than 10 engagements. Hate speakers were exposed to 190,911 popular posts and 685,212 unpopular posts, while non-hate speakers were exposed to 249,194 popular posts and 981,478 unpopular posts. 
While the thresholds are somewhat arbitrary, we considered different thresholds: 1,000 for popular and 20 or 50 for unpopular posts; the results presented below are robust. 
%Engagements are the combined number of likes and retweets a post gets. 
We compute the proportion of low-credibility posts in the two popularity categories, for each user. Fig.~\ref{fig:engagement} shows that when we focus on popular low-credibility posts, the difference between hate and non-hate speakers is not significant ($p = 0.22$). 
However, we observe a significant difference for unpopular low-credibility posts ($p<0.001$). 
This suggests that hate speakers are more likely to be exposed to low-credibility content that is \textit{not} spreading virally. 
While this content may reach fewer users, it is also less likely to be fact-checked. This highlights a potential challenge in fact-checking, emphasizing the need to address misinformation across various levels of online visibility. 

% we need more conclusion
%\textbf{It's a good idea to just combine likes and retweets and call them engagements, since the results are same.}

\subsection{RQ4: Exposure to Political Fake News}

\begin{figure}[t!]
    \centering
    \includegraphics[width=\columnwidth]{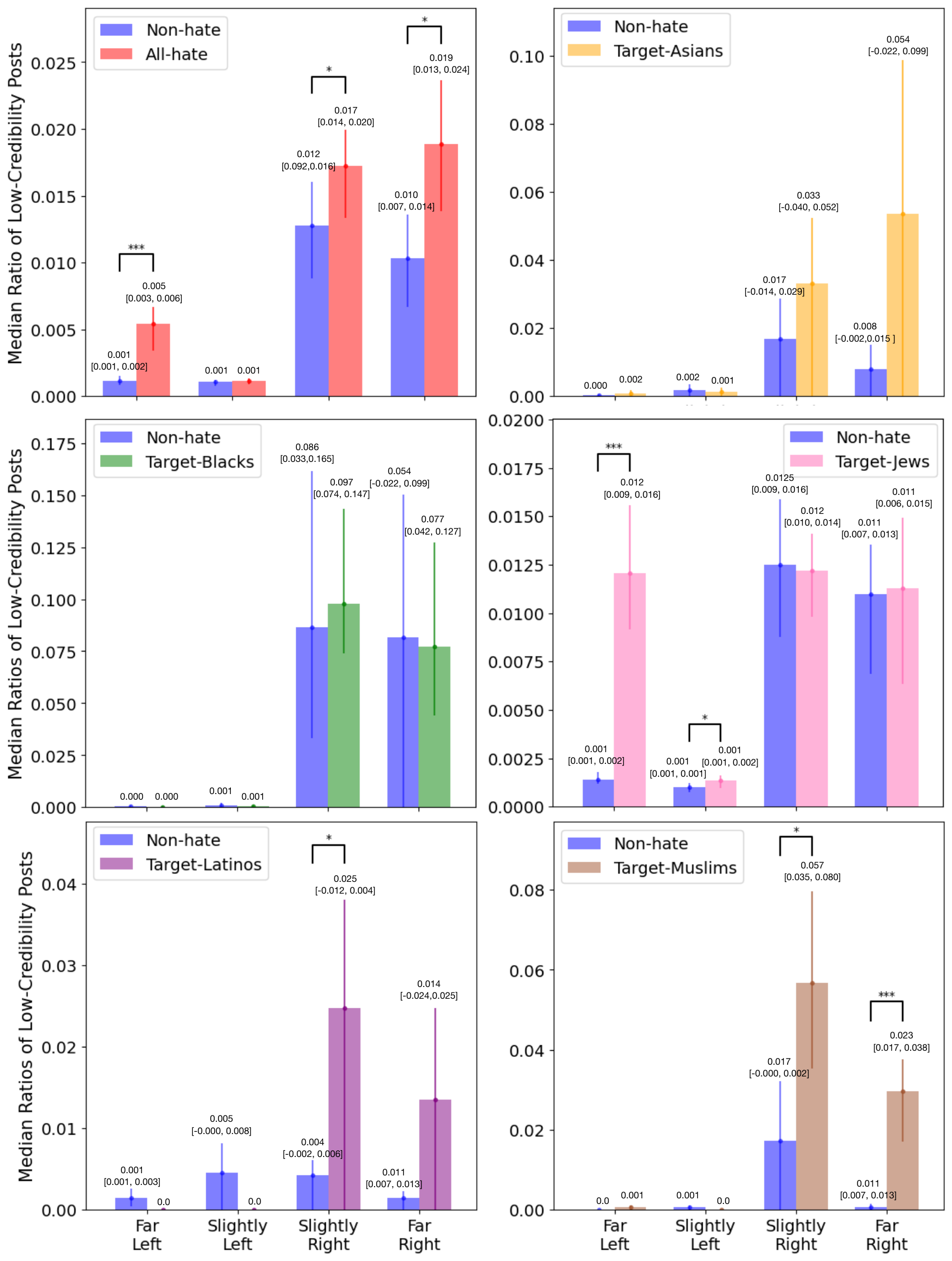}
    \caption{Median proportions of low-credibility news to which hate and non-hate speakers are exposed, broken down by the political alignment of the source for different targets.}
    \label{fig:radical_orientation_by_race}
\end{figure}

% find some references showing tha tht 
% Here, we examined whether the hatful users more encounter radicalized news domain than non-hate speakers. 

% \begin{figure}[h]
%     \centering
%     \includegraphics[width=0.7\linewidth]{Figure/Political_radicalization.png}
%     \caption{The proportion of tweets with different degrees of political orientation in news domain links exposed to hateful and non-hate speakers. }
%     \label{fig:radical_orientation}
% \end{figure}

We finally study the connection between the use of hate speech and exposure to low-credibility news with different political alignment.
%(far left, slightly left, slightly right, and far right). 
Since such a connection with political alignment can depend on the target population of hate speech, we compare hate and non-hate speakers overall as well as for each targeted population.

% We found that tweets containing far-polarized news domains show a significantly higher exposure to hate speakers compared to non-hate speakers. On the other hand, tweets featuring slightly polarized news domains exhibited no statistically significant differences in exposure between hateful and non-hate speakers (see Fig~\ref{fig:radical_orientation}.) \textbf{We are missing the dimension of being left and right and the dimension of low/high credibility.}  

% {Figure/With_outlier/Targets_Political_Orientation_Ratio_absolute_median_credible_news_withoutliers_bytopics.pdf} ==> without text

The median proportions of low-credibility tweets from news sources in four political alignment categories are reported in Fig~\ref{fig:radical_orientation_by_race}. 
Overall, we observe that hate users tend to be exposed to a significantly higher proportion of low-credibility news from far-left ($p<0.001$), slightly right ($p=0.011$), and far-right ($p=0.012$) partisan sources. 
The proportions are higher from conservative than liberal sources for both hate and non-hate speakers. 
Looking at specific target populations, individuals producing anti-semitic content are exposed to significantly higher proportions of low-credibility news from far-left ($p<0.001$) and slightly left ($p = 0.018$) sources. 
Conversely, those engaging in hate speech against Muslims show a significantly higher exposure to slightly right ($p=0.030$) and far-right ($p<0.001$) low-credibility sources. 
We also note higher exposure to slightly right ($p=0.035$) and far-right ($p=0.056$) low-credibility sources for hate speakers targeting Latinos

%The connection between political orientation and the interpretation of fake news from various sources is well-established~\cite{michael2021relationship,axt2020psychological}.

%To ascertain whether the political slant of exposed fake news is also correlated with the engagement in hate speech, we examine the variance in the political orientation of news domains exposed to both hateful and non-hate speakers.

%\begin{figure}[h]
%    \centering
%    \includegraphics[width=0.5\textwidth]{Figure/Political_orientation_race.pdf}
%    \caption{ The proportion of tweets with distinct political orientations exposed to hateful and non-hate speakers}
%    \label{fig:radical_orientation_by_race}
%\end{figure}

\section{Discussion}

In this paper, we study the relationship between the use of hate speech and exposure to fake news on social media. 
We compare the credibility of the posts exposed to 630 hate speakers and 630 non-hate speakers by collecting all the tweets posted by the accounts they follow.   
We find that hate speakers are exposed to significantly higher proportions of tweets linking to low-credibility news sources, which holds for hate speech targeting Muslims and Jews. 
We also find that hate speakers are exposed to higher proportions of unpopular posts linking to low-credibility news sources, while both groups are exposed to roughly the same proportion of popular fake news. 
Finally, anti-semitic speakers are exposed to significantly more posts featuring low-credibility news sources with far-left political alignment, while hate speakers targeting Muslims are exposed to significantly more low-credibility posts from conservative sources. 

Although our study design ensures that exposure to fake news occurs \textit{before} the use of hate speech, we cannot prove a \textit{causal} link from fake news to hate speech. 
There are multiple possible explanations for the observed correlation between hate speech and low-credibility content. 
One is that the higher exposure to misinformation can exacerbate or reinforce existing prejudices and biases, thereby fueling the propagation of hate speech targeting minority groups. In this scenario, the consumption of misinformation acts as a catalyst for the expression of hateful sentiments, creating a feedback loop where exposure to biased or inaccurate information reinforces and amplifies discriminatory attitudes. 
Another possible interpretation of our results is that hate speakers may choose to engage with low-credibility content. Because engagement with users on social media platforms reflects their active interest and preferences, hate speakers may be selectively exposed or drawn to low-credibility news outlets due to their own inclinations or ideological alignment. 

%\paragraph{Limitations:} 
Our analysis has several limitations. 
(i)~The reliance on Twitter data may not represent the diversity of user behaviors across various social media platforms. 
(ii)~Our control group might still include hate speakers, as we did not analyze all of their posts. 
(iii)~We primarily focus on racism-related hate speech due to the data availability. Hate speech, however, encompasses diverse forms of discrimination based on other attributes. Further analysis is needed to generalize our results to all forms of hate speech. 
(iv)~Our use of content posted by followed users as a proxy for exposure has limitations: on the one hand, not all content posted by followed accounts may be seen by users, and on the other hand, users may be exposed to content via other mechanisms, such as search and recommendations. While reconstructing actual news feeds would have been ideal, it was unfortunately impossible even when the Twitter API was still available to researchers.   
Finally, (v)~our fake news analysis relies on credibility labels at the source level. Given that low-credibility news outlets can occasionally produce accurate content, additional analysis at the individual article level could provide a more comprehensive understanding of fake news. 

Despite such limitations, this study contributes to our understanding of the interplay between hate speech and exposure to fake news, which can help inform interventions aiming to mitigate either hate speech or harmful misinformation.
The finding that hate speakers are exposed to lower volumes of content from credible news sources emphasizes the importance of promoting reliable information, providing alternatives to interventions that focus solely on the prevalence of misinformation.

\paragraph{Ethical Impact.}
Due to privacy concerns, we limited our analysis to the tweets public profiles posted and were exposed to, avoiding the use of identifiable information. We only reported aggregate results. We also do not share the data due to privacy and changes in X/Twitter data collection and sharing policies. We acknowledge the sensitivity of the topic under study and strive to maintain inclusive language to the best of our ability. A potential negative impact of this study may be the alienation of people who used hate speech by labeling them as hate speakers. Our objective is not to target those people but to understand the factors associated with their harmful behavior, in order to mitigate such behavior. We emphasize that unsupported conclusions, such as claiming that anyone who is exposed to fake news is a hate speaker, are not in line with our research. We discourage any such misuse of our work.

%I don't think this is noteworthy, so commenting it
%It is also noteworthy that the sample size of hate speakers is disproportionately concentrated, particularly in instances of hate speech targeting Jews. 

\paragraph{Acknowledgements.}
This work was supported in part by the Knight Foundation.

\bibliography{aaai22}

%\subsection{Detailed Instructions}
%Please do not modify the questions and only use the provided macros for your answers. In your paper, please delete all text in the \textbf{Overview} and \textbf{Detailed Instructions} sections, as well as all subsection headers, keeping only the \textbf{Checklist} section heading above along with the questions/answers below. 

%For each question, change the default \answerTODO{Answer} to \answerYes{Yes, and},
%\answerNo{No, because}, or \answerNA{NA}, when the question seems inappropriate for your research study. You are strongly encouraged to include a {\bf justification to your answer}, either by referencing the appropriate section of your paper or providing a brief inline description. Within the Checklist section too, you may supplement your answers with a brief discussion that expands on answers to the checklist where necessary. For example:
%\begin{itemize}
%  \item Did you include the license to the code and datasets? \answerYes{Yes, see the Methods and the Appendix.}
%  \item Did you include the license to the code and datasets? \answerNo{No, because the code and the data are proprietary.}
%  \item Did you include the license to the code and datasets? \answerNA{NA}
%\end{itemize}
%%% END INSTRUCTIONS %%%

\subsection{Ethics Checklist}

\begin{enumerate}

\item For most authors...
\begin{enumerate}
    \item  Would answering this research question advance science without violating social contracts, such as violating privacy norms, perpetuating unfair profiling, exacerbating the socio-economic divide, or implying disrespect to societies or cultures?
    \answerYes{Yes, please see Introduction, Discussion and Ethical Impact}
  \item Do your main claims in the abstract and introduction accurately reflect the paper's contributions and scope?
    \answerYes{Yes}
   \item Do you clarify how the proposed methodological approach is appropriate for the claims made? 
    \answerYes{Yes, please see Introduction}
   \item Do you clarify what are possible artifacts in the data used, given population-specific distributions?
    \answerYes{Yes, we explicitly state that the distribution of user groups targeting different populations is skewed towards the group targeting Jews in the Data and Methodology section. So we break down our analysis to different target populations in RQ2 and RQ4.}
  \item Did you describe the limitations of your work?
    \answerYes{Yes, please see Limitations at the end of Discussion}
  \item Did you discuss any potential negative societal impacts of your work?
    \answerYes{Yes, please see Ethical Impact section}
      \item Did you discuss any potential misuse of your work?
    \answerYes{Please see Ethical Impact Section}
    \item Did you describe steps taken to prevent or mitigate potential negative outcomes of the research, such as data and model documentation, data anonymization, responsible release, access control, and the reproducibility of findings?
    \answerYes{Yes, we opted out of sharing the data publicly and presenting only aggregate results as we discussed in Ethical Impact Section}
  \item Have you read the ethics review guidelines and ensured that your paper conforms to them?
    \answerYes{Yes}
\end{enumerate}

\item Additionally, if your study involves hypotheses testing...
\begin{enumerate}
  \item Did you clearly state the assumptions underlying all theoretical results?
    \answerNA{NA}
  \item Have you provided justifications for all theoretical results?
    \answerNA{NA}
  \item Did you discuss competing hypotheses or theories that might challenge or complement your theoretical results?
    \answerNA{NA}
  \item Have you considered alternative mechanisms or explanations that might account for the same outcomes observed in your study?
   \answerNA{NA}
  \item Did you address potential biases or limitations in your theoretical framework?
    \answerNA{NA}
  \item Have you related your theoretical results to the existing literature in social science?
   \answerNA{NA}
  \item Did you discuss the implications of your theoretical results for policy, practice, or further research in the social science domain?
    \answerNA{NA}
\end{enumerate}

\item Additionally, if you are including theoretical proofs...
\begin{enumerate}
  \item Did you state the full set of assumptions of all theoretical results?
    \answerNA{NA}
	\item Did you include complete proofs of all theoretical results?
    \answerNA{NA}
\end{enumerate}

\item Additionally, if you ran machine learning experiments...
\begin{enumerate}
  \item Did you include the code, data, and instructions needed to reproduce the main experimental results (either in the supplemental material or as a URL)?
    \answerNA{NA}
  \item Did you specify all the training details (e.g., data splits, hyperparameters, how they were chosen)?
    \answerNA{NA}
     \item Did you report error bars (e.g., with respect to the random seed after running experiments multiple times)?
    \answerNA{NA}
	\item Did you include the total amount of compute and the type of resources used (e.g., type of GPUs, internal cluster, or cloud provider)?
    \answerNA{NA}
     \item Do you justify how the proposed evaluation is sufficient and appropriate to the claims made? 
    \answerNA{NA}
     \item Do you discuss what is ``the cost`` of misclassification and fault (in)tolerance?
    \answerNA{NA}
  
\end{enumerate}

\item Additionally, if you are using existing assets (e.g., code, data, models) or curating/releasing new assets...
\begin{enumerate}
  \item If your work uses existing assets, did you cite the creators?
    \answerYes{Yes, we used a preexisting dataset and cited the creators}
  \item Did you mention the license of the assets?
    \answerNo{No, the dataset was shared by the author}
  \item Did you include any new assets in the supplemental material or as a URL?
    \answerNo{No}
  \item Did you discuss whether and how consent was obtained from people whose data you're using/curating?
    \answerNo{No, the data only consist of public posts}
  \item Did you discuss whether the data you are using/curating contains personally identifiable information or offensive content?
    \answerYes{Yes, but we are not using user data, only the posts they authored or exposed to, which we state in Ethical Impact.}
\item If you are curating or releasing new datasets, did you discuss how you intend to make your datasets FAIR?
\answerNA{NA}
\item If you are curating or releasing new datasets, did you create a Datasheet for the Dataset? 
\answerNA{NA}
\end{enumerate}

\item Additionally, if you used crowdsourcing or conducted research with human subjects...
\begin{enumerate}
  \item Did you include the full text of instructions given to participants and screenshots?
    \answerNA{NA}
  \item Did you describe any potential participant risks, with mentions of Institutional Review Board (IRB) approvals?
    \answerNA{NA}
  \item Did you include the estimated hourly wage paid to participants and the total amount spent on participant compensation?
   \answerNA{NA}
   \item Did you discuss how data is stored, shared, and deidentified?
   \answerNA{NA}
\end{enumerate}

\end{enumerate}

\end{document}